\input{aipcheck}
\documentclass[
    ,final             
  ]
  {aipproc}
\layoutstyle{6x9}
\usepackage{epsfig}
\begin{document}
\def\be{\begin{equation}}
\def\ee{\end{equation}}
\def\ba{\begin{eqnarray}}
\def\ea{\end{eqnarray}}
\def\Mesz{M\'esz\'aros}
\def\siml{\lower4pt \hbox{$\buildrel < \over \sim$}}
\def\simg{\lower4pt \hbox{$\buildrel > \over \sim$}}
\def\etal{{\it et al.}}
\def\msun{M_\odot}
\def\eps{\epsilon}
\newcommand{\figuresize}{0.41\textwidth}
\newcommand{\boxsize}{0.89\textwidth}
\newcommand{\smallboxsize}{0.8\textwidth}
\title{Early afterglow, magnetized central engine, 
       and a quasi-universal jet configuration for long GRBs}


\author{Bing Zhang$^{1}$, Shiho Kobayashi$^{1,2}$, 
Peter M\'esz\'aros$^{1,2,3}$, \\
Nicole M. Lloyd-Ronning$^{4}$ and Xinyu Dai$^{1}$ }
{
  address={
$^1$Department of Astronomy \& Astrophysics, 
Penn State University,
University Park, PA 16802, USA, \\
$^2$Department of Physics,
Penn State University,
University Park, PA 16802, USA\\
$^3$The Institute for Advanced Study, Princeton, NJ 08540, USA \\ 
$^4$Los Alamos National Laboratory, 
MS B244, Los Alamos, NM 87544, USA\\}
}
\begin{abstract}
Two separate topics are discussed. (1) We describe the classifications
of the long GRB early afterglow lightcurves within the framework of
the fireball shock model, focusing on the interplay between the
reverse and forward shock emission components. We will also provide
evidence that the central engine of at least two bursts are entrained
with strong magnetic fields, and discuss the implications of this
result for our understanding of the GRB phenomenon; (2) We argue that
the current gamma-ray burst (GRB) and X-ray flash (XRF) data are
consistent with a picture that all GRB-XRF jets are structured and
quasi-universal, with a typical Gaussian-like jet structure. 
\end{abstract}
\maketitle

\section{Early afterglows}

\subsection{Classifications}

A GRB fireball is eventually decelerated by an ambient medium. During
the deceleration, a long-lived forward shock propagates into the
medium, and a short-lived reverse shock propagates into the
fireball shell\cite{mr97}. The former is responsible for the long-term
afterglow emission, while the latter contributes a noticeable emission
component at the very early afterglow epoch\cite{mr97,sp99,k00}. 
So a GRB early afterglow is the interplay between the reverse and the
forward shock emission components, and its diagnose would reveal rich
information about the GRB fireball and the ambient medium. Very
early optical afterglows have now been detected for a handful of
GRBs\cite{ak99,fox03,li03}, and the {\em Swift} GRB mission, scheduled 
to be launched in June 2004, will greatly increase the sample of the
GRB early afterglow data. Here we discuss the classifications of the
early afterglow lightcurves within the framework of the fireball shock
model. The predictions will be fully confronted by the future abundant 
early afterglow data.

\begin{figure}[htb]
 \begin{minipage}[t]{0.35 \textwidth}
 \epsfxsize=\boxsize
 \epsfbox{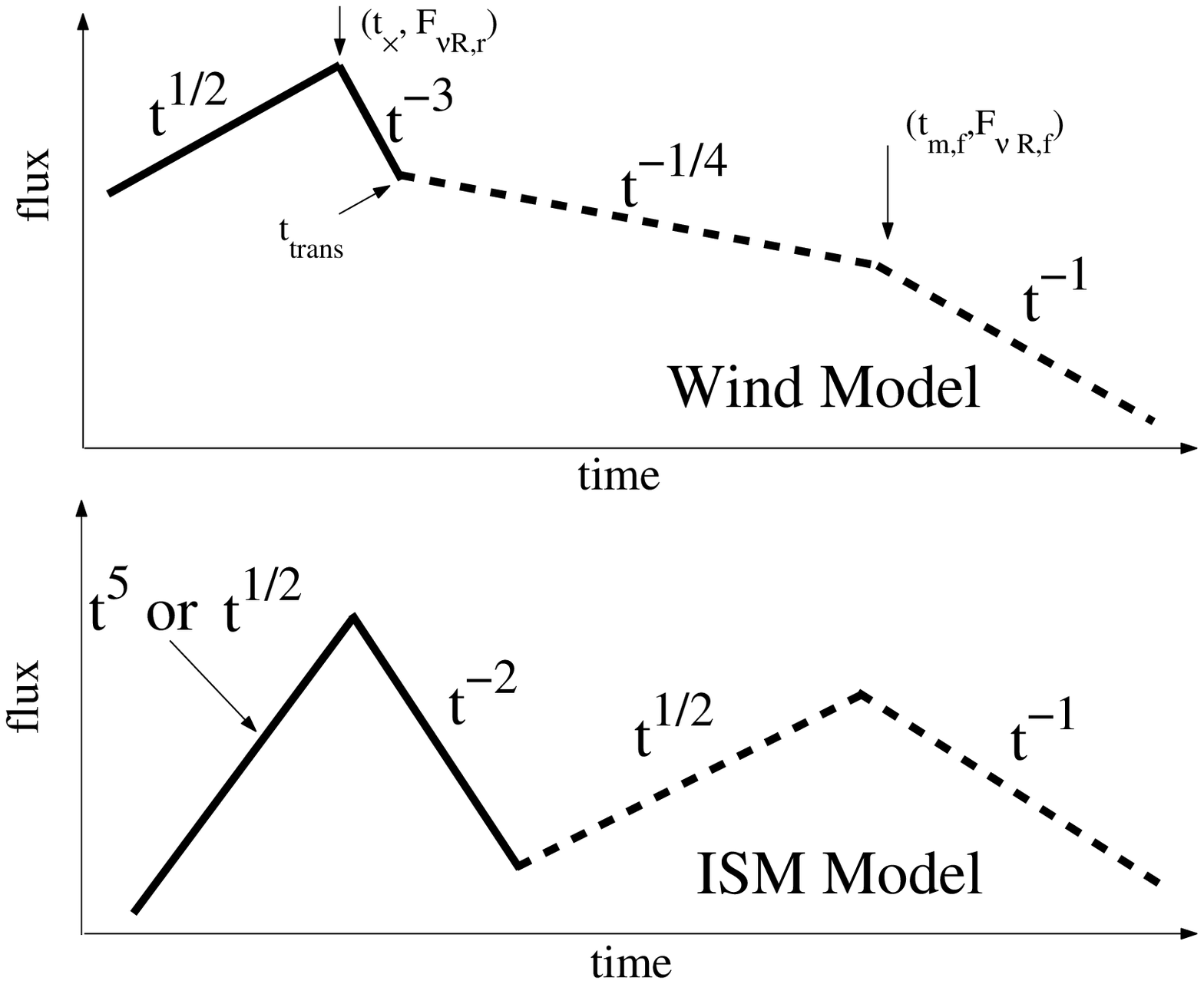}
 \end{minipage}
 \hspace{3.5mm}
 \begin{minipage}[t]{0.35 \textwidth}
 \epsfxsize=\boxsize
 \epsfbox{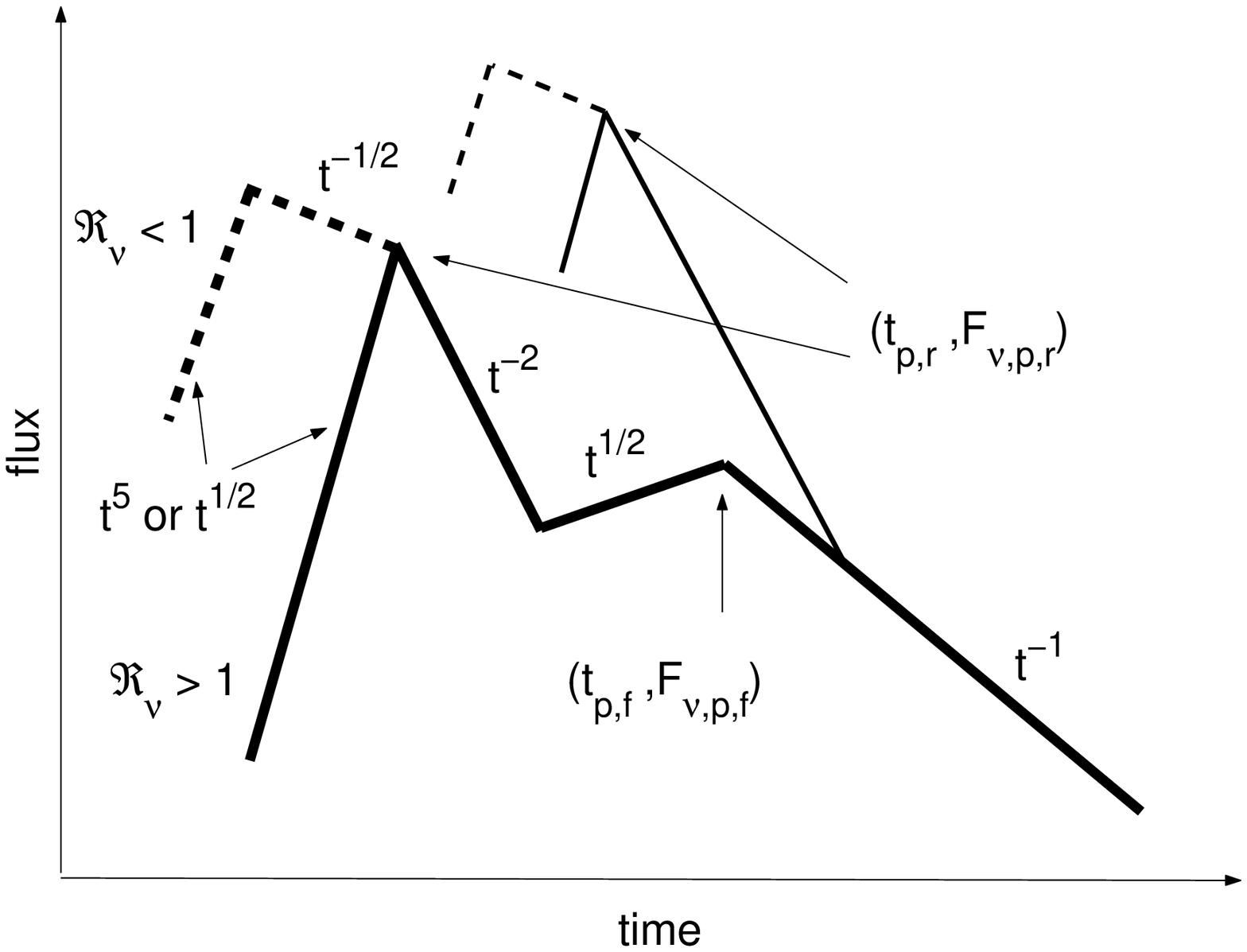}
 \end{minipage}
 \hspace{3.5mm}
 \begin{minipage}[t]{0.35\textwidth}
 \epsfxsize=\boxsize
 \epsfbox{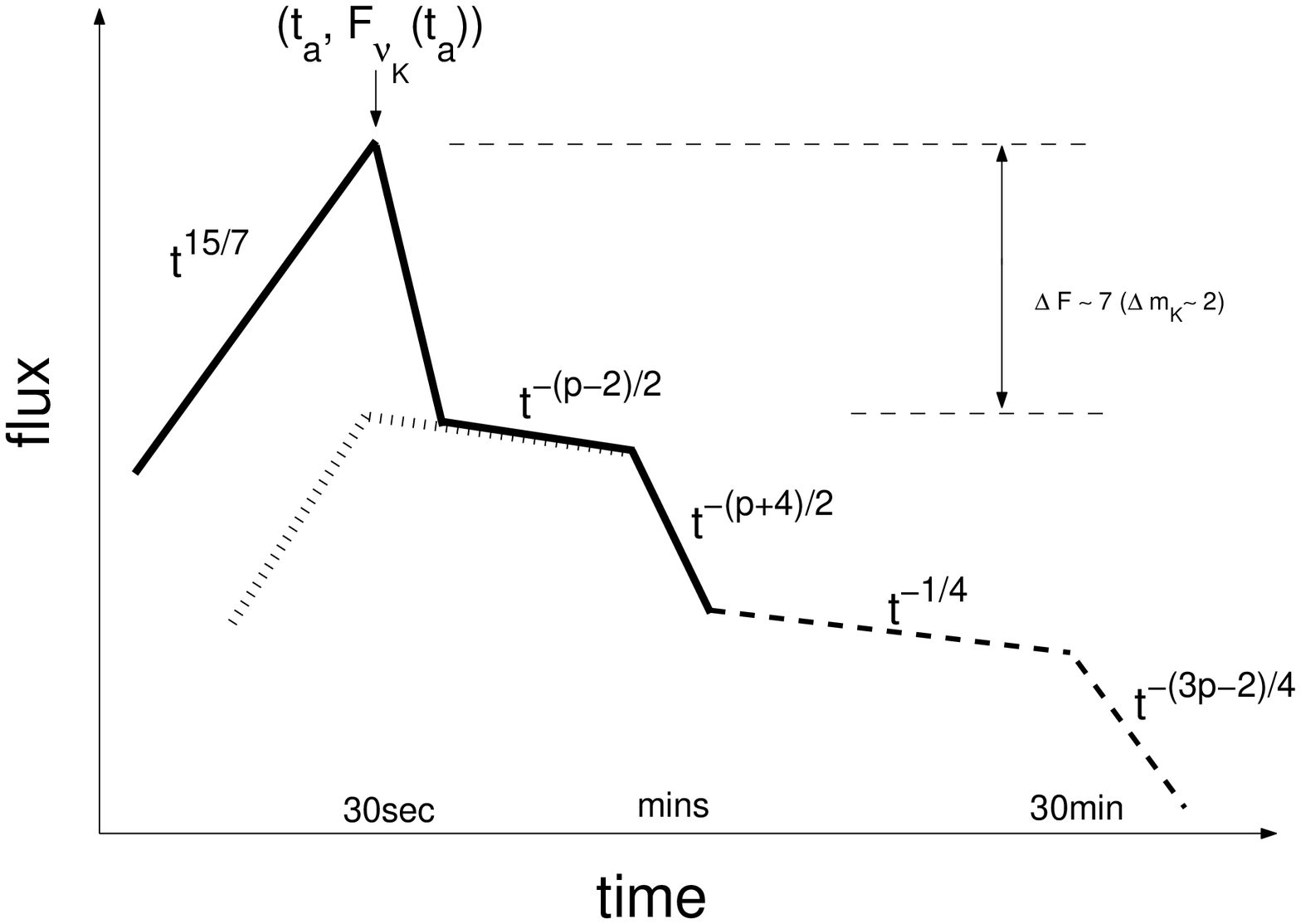}
 \end{minipage}
 \caption{{\it Left panel}: Typical early optical afterglow
 lightcurves for the ISM (bottom) and the wind (top)
 cases\cite{kz03b}. Solid lines are for the reverse shock component,
 and the dashed lines are for the foreward shock emssion component;
 {\it Middle Panel}: Two types early optical afterglow lightcurves
 for the ISM case\cite{zkm03}. Type I (thick line) is the
 rebrightening type, with a distinct separation of the reverse and
 forward shock components. Type II (thin line) is the flattening
 type, with the forward shock peak burried beneath the reverse shock
 component. A Type II lightcurve is usually associated with a strongly 
 magnetized fireball;
 {\it Right panel}: Another typical optical/IR lightcurve for the wind 
 case\cite{kmz03}. This is applicable when the inverse Compton
 scattering is not important in the reverse shock region.}. 
 \label{fig:lightcurves}
 \end{figure}

In general, the early afterglow lightcurves can be categorized
according to the type of the ambient medium. Two well-discussed types
of medium include a constant density medium ($n=$const) which is
applicable for interstellar medium (ISM), and a wind-type medium
($n\propto r^{-2}$), which is typical for the environment of a
pre-burst massive star progenitor. The left panel of Figure
\ref{fig:lightcurves} outlines 
the typical optical early afterglow lightcurves for ISM (bottom) and
wind (top) cases, respectively\cite{kz03b}. In both cases, the reverse 
shock emission component peaks at the time when the reverse shock
crosses the shell, and drops rapidly after the peak. For the forward
shock component, there is a peak for the ISM case corresponding to the 
crossing of the typical synchrotron frequency across the
band\cite{spn98,kz03a}, while for the wind case, the flux fades
exclusively\cite{cl99}.

For the ISM case, the early afterglow lightcurves could be further
categorized into two types\cite{zkm03} (see the middle panel of
Fig. \ref{fig:lightcurves}). For typical parameters and assuming that 
the shock parameters (equipartition parameters $\epsilon_e$ and
$\epsilon_B$, as well as the power-law index of the particle
distribution $p$) are the same in both shocks, the lightcurve (Type I)
is characterized by a ``re-brightening'' feature, i.e., there are
two distinct lightcurve peaks for both shocks. Conversely, under
certain conditions, the forward shock peak is burried beneath the
reverse shock emission component, and the lightcurve (Type II) is
characterized by a ``flattening'' feature. A Type-II lightcurve
usually requires a stronger magnetic field in the reverse shock region 
than in the forward shock region, i.e., refers to a strongly
magnetized central engine.

For the wind case, the early afterglow lightcurves could be also
further categorized into two types based on whether synchrotron
self-inverse Compton (IC) emission is important. A wind-type 
medium implies a low cooling frequency and a high self-absorption
frequency. The synchrotron self-absorption effect prevents the
electrons from cooling, so that electrons are potentially piled up
near the self-absorption energy\cite{kmz03}. If the IC effect is
important, this additional cooling mechanism tends to destroy the
pile-up bump, so that the treatment that neglects it gives the
(approximately) correct description of the lightcurves\cite{kz03b}
(top lightcurve, left panel, Fig. \ref{fig:lightcurves}). If the
IC cooling is less important compared with the synchrotron cooling, as
is expected for a strongly magnetized central engine\cite{zkm03},
the electron pile-up effect is prominent, which implies a
bump in the synchrotron spectrum and hence, another bump in the early
afterglow lightcurve\cite{kmz03} (see right panel,
Fig. \ref{fig:lightcurves}). Detections of such a bump would provide
valuable information to estimate the wind mass lose and other fireball 
parameters\cite{kmz03}.

\subsection{A Strongly Magnetized GRB Central Engine}

Early afterglow lightcurves could be used to constrain important
fireball parameters, such as the initial Lorentz factor, wind mass
loss, etc\cite{zkm03,kz03b,kmz03}. Another important piece of
information is about the magnetic content of the fireball. A strongly
magnetized central engine is widely speculated to power GRBs on many
grounds (e.g. \cite{zm03} for a review). If this is the case, the
magnetic field in the reverse shock region is expected to be stronger
than that in the forward shock region, since the fireball itself would 
carry some fields from the central engine. Defining $R_B=B_r/B_f$ as a 
free parameter (where $B_r$ and $B_f$ are the magnetic field
strengths in the reverse shock and forward shock, respectively), one
can use the early afterglow lightcurves to constrain $R_B$\cite{zkm03}. 
Using a straightforward analysis by combining both the reverse and the 
forward shock emission data, we have performed detailed case studies
for the GRBs that have early afterglow detections. The results suggest
that $R_B$ is larger than unity for both GRB 990123 and GRB
021211\cite{zkm03}. The latter result is confirmed by a more detailed,
independent study\cite{kp03}. These results suggest that at least for
some bursts, the central engine is likely entrained with strong
magnetic fields. The discovery of strong linear polarization\cite{cb03} 
of gamma-ray emission in GRB 021206 is also consistent with such a
picture. 

An important question is how strong the magnetic energy density 
is as compared with the kinetic energy density. 
Conventionally one can define $\sigma=L_P/L_K$ to categorize the
fireball (where $L_P$ and $L_K$ are the Poynting flux luminosity and
kinetic energy luminosity, respectively). The canonical fireball is in 
the $\sigma \ll 1$ regime. For GRB 990123, broad-band modeling
suggests that $\epsilon_{B,f} \sim 7.4\times 10^{-4}$\cite{pk02}. Our
analysis indicates that $R_B=(\epsilon_{B,r}/\epsilon_{B,f}) \sim 15$
for this burst, so that $\epsilon_{B,r} \sim 0.17$. This is still in
the $\sigma \ll 1$ regime, which ensures self-consistency of our
hydrodynamical treatment. In the meantime, it suggests that the
magnetitized fireball is not Poynting-flux dominated at the
deceleration radius. 

\section{A quasi-universal structured jet model}

\subsection{Uniform vs. Universal Jets}

One intriguing finding in the GRB afterglow observations is that the
geometry-corrected total energy for different bursts is
standard\cite{f01,bfk03}. There are two equivalent
interpretations. One is that different GRBs collimate a same amount of 
energy in different solid angles, but with a uniform energy
distribution within the jets. The other is that all GRBs have a same
jet configuration, but the energy per solid angle decreases with angle
from the jet axis in the form of $\epsilon(\theta) \propto
\theta^{-2}$, so that the inferred jet angles from the afterglow data
correspond to the observers' viewing angles\cite{rlr02,zm02a}. The
former is called ``uniform jets'', and the latter is called
``universal jets''. 

These two models are two extremal presentations of what might happen
in reality. In realistic simulations such as those in the collapsar
model, the emerging GRB jets natually have a non-uniform angular
structure\cite{zwm03}. On the other hand, it is unrealistic to expect
that all GRB jets are exactly universal. Such an exactly universal
picture is already disfavored by the $\log(E_{iso})-\log(\theta_j)$
plot of the observed data, which indicate a large scatter around the
$E_{iso}\propto \theta_j^{-2}$ line (see solid squares in
Fig. \ref{fig:eiso-thetaj}).

\subsection{Quasi-Universal Jets: Power Law vs. Gaussian}

A reasonable picture is that GRB jets preserve certain angular
structure individually, and may have a ``quasi-universal'' pattern of
the jet structure\cite{ldz03,zdlm03}. The so-called quasi-universal
jet model suggests that all GRBs have a more-or-less similar angular
jet structure, with the model parameters (e.g. the power-law index for
the power-law jets, the typical angle for the Gaussian jets, and the
normalization parameters for both types of jets) being distributed
around some standard values with a small scatter.

 \begin{figure}[htb]
 \begin{minipage}[t]{0.4 \textwidth}
 \epsfxsize=\boxsize
 \epsfysize=1.5in
 \epsfbox{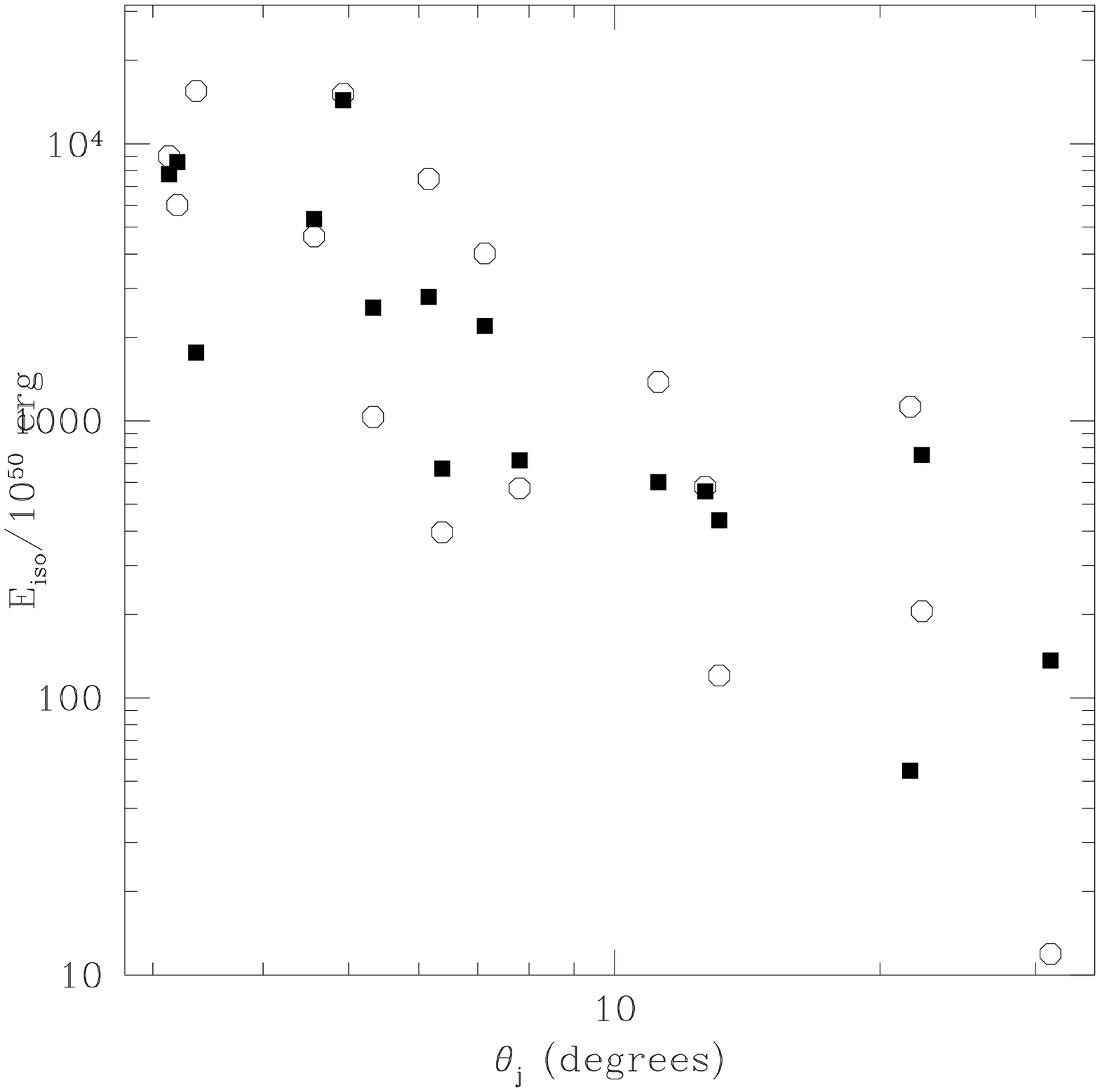}
 \end{minipage}
 \hspace{3mm}
 \begin{minipage}[t]{0.4\textwidth}
 \epsfxsize=\boxsize
 \epsfysize=1.5in
 \epsfbox{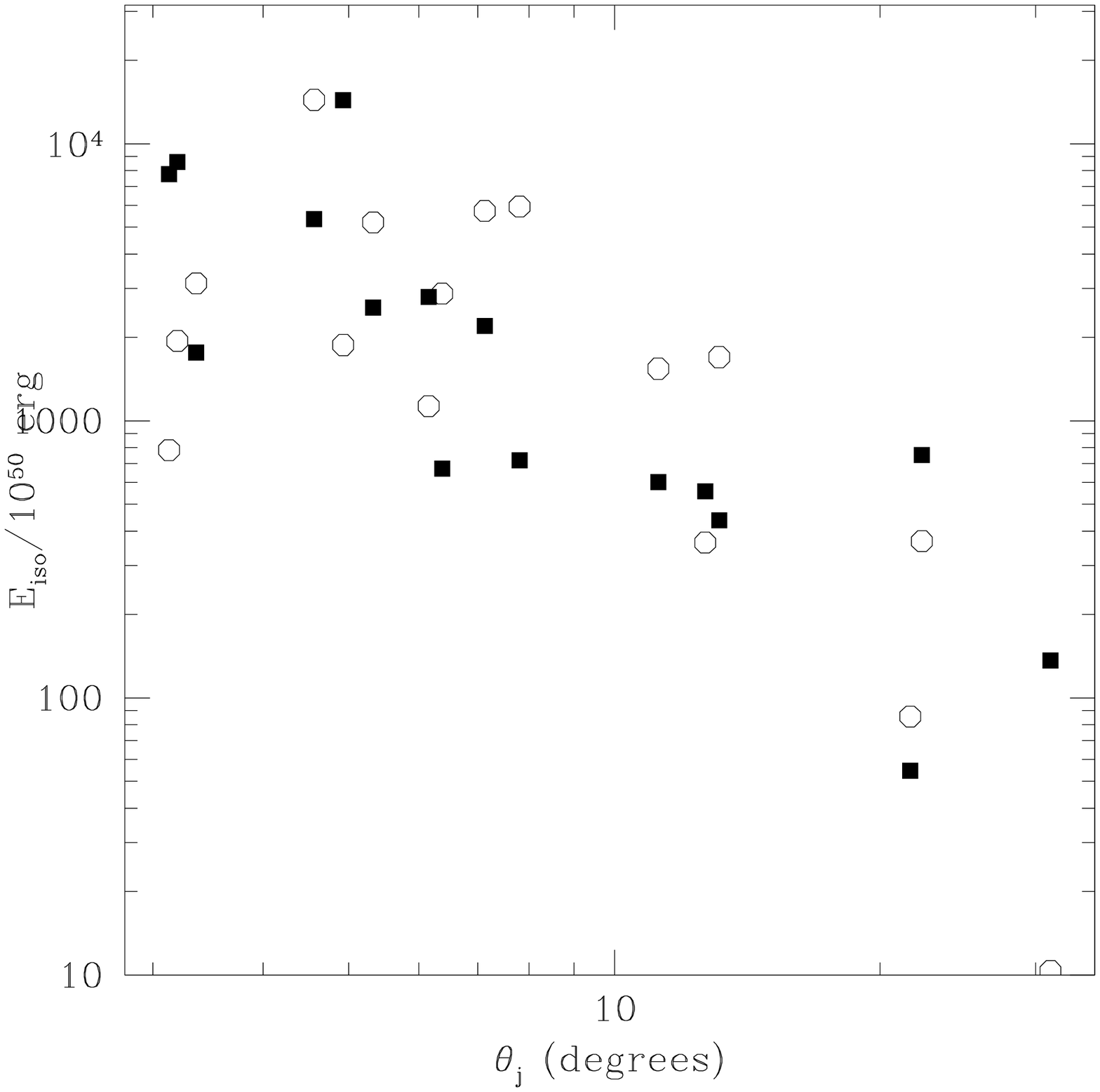}
 \end{minipage}
 \caption{Simulated $E_{iso}-\theta_j$ data from the quasi-universal 
 structured jet models\cite{ldz03} (open circles) as compared with the 
 data\cite{bfk03} (solid squares). 
 {\it Left panel:} Quasi-universal power-law model;
 {\it Right panel}: Quasi-universal Gaussian model.}
  \label{fig:eiso-thetaj}
 \end{figure}

When parameters are allowed to have some scatter, the $k=-2$ power law
structure is no longer a pre-requisite for individual bursts. Other
types of jet structure (such as Gaussian)\cite{zm02a} are also
allowed, especially when the total energy within the jet is preserved
to be a quasi-constant. Figure \ref{fig:eiso-thetaj} shows that both a
quasi-universal power-law model and a quasi-universal Gaussian model
can reproduce the $E_{iso}-\theta_j$ data\cite{ldz03}.

\subsection{Quasi-Universal Gaussian Jets: A Unified Model for GRBs
and XRFs}

X-ray flashes (XRFs) are the natural extension of GRBs towards the
softer and fainter regime. Recent HETE-2 data reveal that an
intriguing empirical relation $E_p \propto (E_{iso})^{1/2}$ (where
$E_p$ is the peak energy of the GRB-XRF spectrum)\cite{amati02} is
extended from GRBs to XRFs\cite{lamb03}, and that the number
ratio among GRBs, X-ray rich GRBs (XRGRBs) and XRFs is roughly
1:1:1. These facts pose severe constraints on both the
universal\cite{lamb03} and the uniform\cite{zdlm03} jet models. The
current GRB-XRF prompt emission and afterglow data are, however, 
consistent with a quasi-universal Gaussian jet model\cite{zdlm03}.

 \begin{figure}[htb]
 \begin{minipage}[t]{0.35 \textwidth}
 \epsfxsize=\boxsize
 \epsfysize=1.5in
 \epsfbox{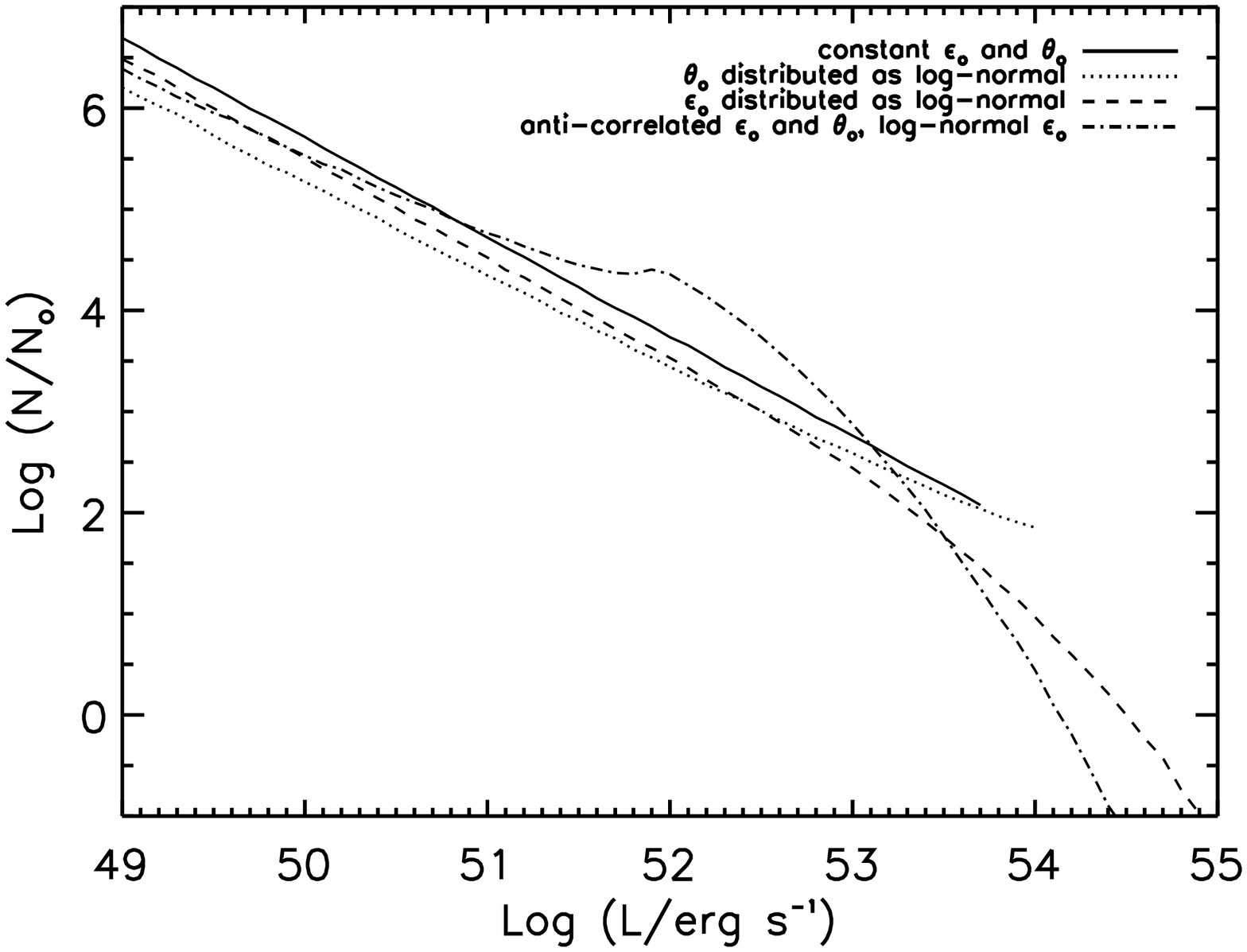}
 \end{minipage}
 \hspace{3mm}
 \begin{minipage}[t]{0.35\textwidth}
 \epsfxsize=\boxsize
 \epsfysize=1.5in
 \epsfbox{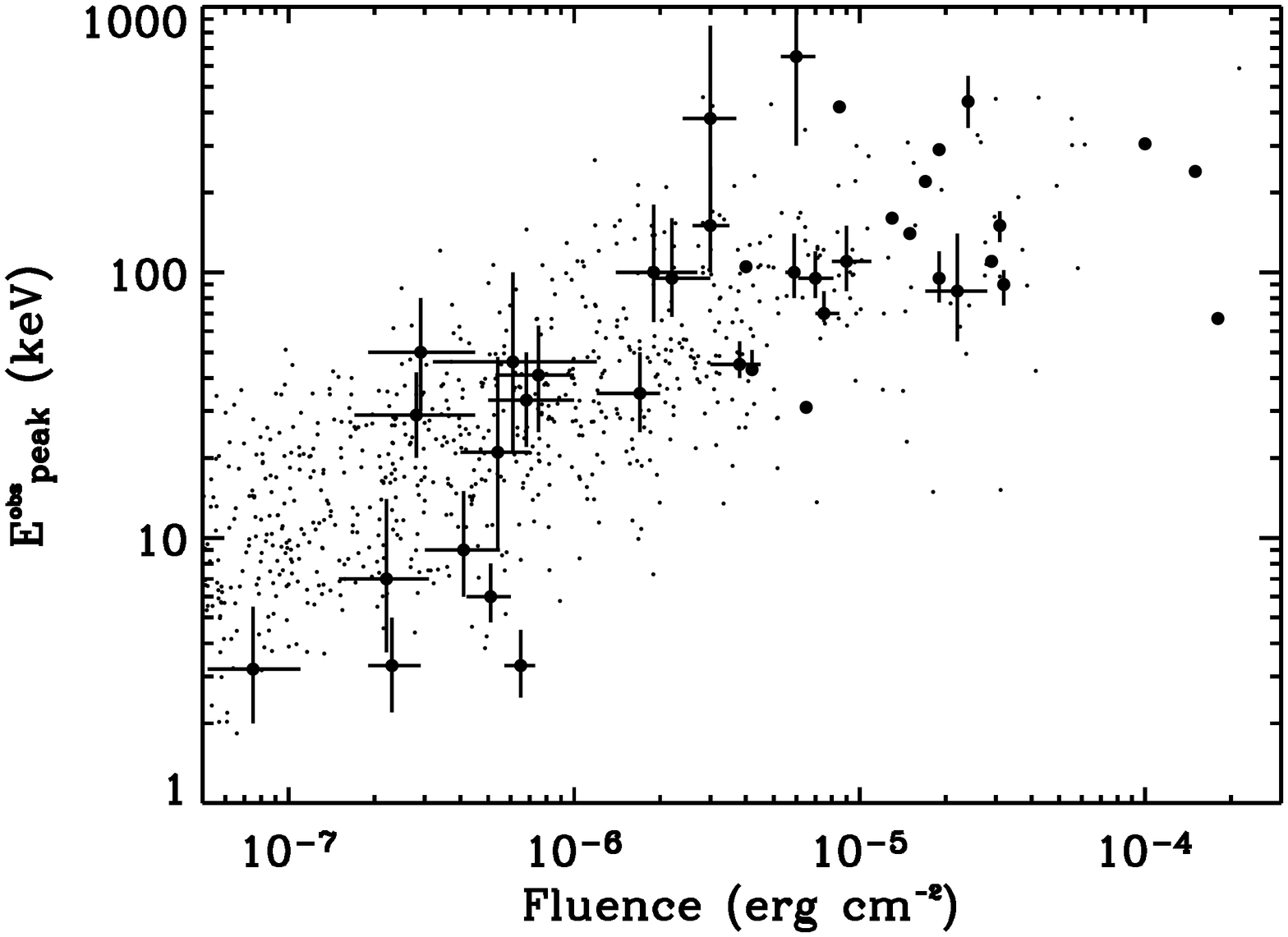}
 \end{minipage}
 \hspace{3mm}
 \begin{minipage}[t]{0.35\textwidth}
 \epsfxsize=\boxsize
 \epsfysize=1.5in
 \epsfbox{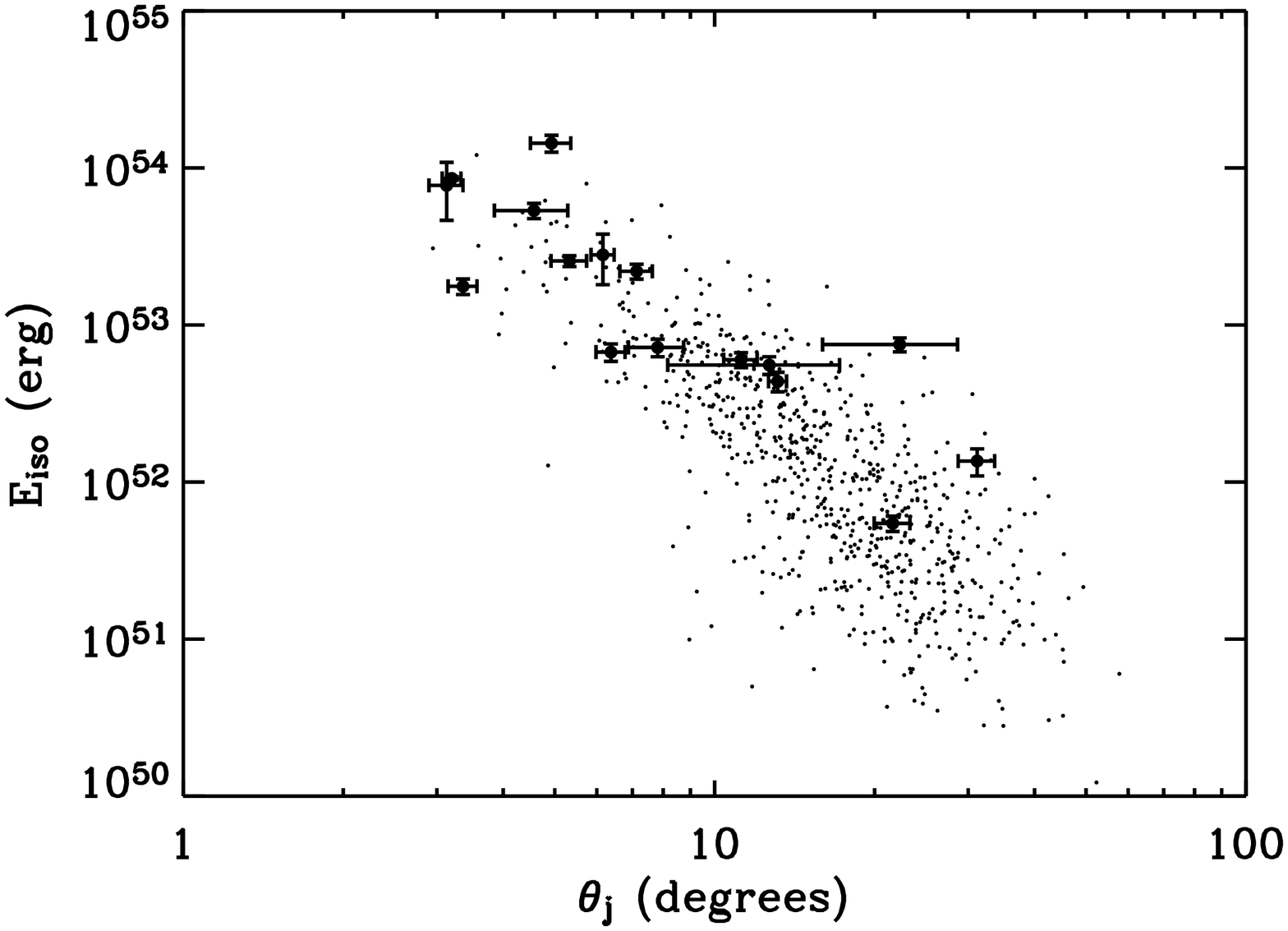}
 \end{minipage}
 \caption{A quasi-universal Gaussian jet model confronted with the
current data. {\it Left:} the predicted GRB luminosity
functions\cite{ldz03}; {\it Middle:} the $E_p$-fluence
diagram\cite{zdlm03}; {\it Right:} the $E_{iso}-\theta_j$
diagram\cite{zdlm03}.} 
 \label{fig:quasi}
 \end{figure}

Figure \ref{fig:quasi} presents some predictions of the
quasi-universal Gaussian jet model. The GRB luminosity function (left
panel) is predicted to be a broken power-law with indices changing
from -1 to $\sim -2$\cite{ldz03}. This is consistent with some
luminosity function studies. The GRB:XRGRB:XRF number ratio is roughly
1:1:1 (middle panel), and the afterglow $E_{iso}-\theta_j$ correlation
is consistent with the ``standard energy reservoir''
relation\cite{zdlm03}. More rigorous tests of this model with a
wider spectrum of data are being performed\cite{dz04}.


\begin{theacknowledgments}
This research is supported partly through NASA NAG5-13286, 
NSF PHY 01-14375, and the Monell Foundation.
\end{theacknowledgments}


\bibliographystyle{aipproc}   


\IfFileExists{\jobname.bbl}{}
 {\typeout{} \typeout{******************************************}
  \typeout{** Please run "bibtex \jobname" to optain} \typeout{** the
  bibliography and then re-run LaTeX} \typeout{** twice to fix the
  references!}  \typeout{******************************************}
  \typeout{} }

\end{document}